\documentclass[11pt]{article}
\pdfoutput=1
\usepackage{arxiv}
\usepackage{microtype}
\usepackage[table]{xcolor}
\usepackage[T1]{fontenc}
\usepackage{amsthm}
\usepackage{amssymb}
\usepackage[english]{babel}
\usepackage[utf8]{inputenc}
\usepackage{amsmath}
\usepackage{amsfonts}
\usepackage{graphicx}
\usepackage[colorinlistoftodos]{todonotes}
\usepackage{mathtools}
\usepackage{enumitem}
\usepackage{thmtools}
\usepackage{fullpage}
\usepackage{booktabs}
\usepackage{bbm}
\usepackage[normalem]{ulem} 
\usepackage[hidelinks]{hyperref}
\usepackage{multirow}
\newcommand{\A}{\mathbf{A}}

\newcommand{\vol}{\textrm{vol}}
\newcommand{\ABCD}{\textbf{ABCD}}
\newcommand{\hABCD}{\textbf{h--ABCD}}

\theoremstyle{plain}

\newcommand{\cd}{\text{cd}}
\newcommand{\kl}{\text{kl}}

\title{Predicting Properties of Nodes via Community-Aware Features}

\author{
Bogumi\l{} Kami\'nski\thanks{Decision Analysis and Support Unit, SGH Warsaw School of Economics, Warsaw, Poland; e-mail: \texttt{bkamins@sgh.waw.pl}}
\And
Pawe\l{}~Pra\l{}at\thanks{Department of Mathematics, Toronto Metropolitan University, Toronto, ON, Canada; e-mail: \texttt{pralat@torontomu.ca}}
\And
Fran\c{c}ois Th\'eberge\thanks{Tutte Institute for Mathematics and Computing, Ottawa, ON, Canada; email: \texttt{theberge@ieee.org}}
\And
Sebastian Zaj\k{a}c\thanks{Decision Analysis and Support Unit, SGH Warsaw School of Economics, Warsaw, Poland; e-mail: \texttt{szajac2@sgh.waw.pl}}
}

\begin{document}

\maketitle

\begin{abstract}
\noindent
This paper shows how information about the network's community structure can be used to define node features with high predictive power for classification tasks. To do so, we define a family of community-aware node features and investigate their properties. Those features are designed to ensure that they can be efficiently computed even for large graphs. We show that community-aware node features contain information that cannot be completely recovered by classical node features or node embeddings (both classical and structural) and bring value in node classification tasks. This is verified for various classification tasks on synthetic and real-life networks.

\bigskip

\end{abstract}

\section*{Statements and Declarations}
BK and SZ have been supported by the Polish National Agency for Academic Exchange under the Strategic Partnerships programme, grant number BPI/PST/2021/1/00069/U/00001.

\newpage

\section{Introduction}

Classification is a classical supervised machine learning problem in which data instances with ground truth labels are used to train a model that can predict the labels of unseen data instances. In the context of data that can be represented as graphs, node classification is a particularly important problem in which the goal is to predict labels associated with the nodes. Node classification is widely used in various practical applications such as social network analysis~\cite{bhagat2011node}, recommender systems~\cite{ying2018graph}, and applied chemistry~\cite{gilmer2017neural}.

Many node classification methods have been previously investigated in the literature, such as generalized (regularized) linear classifiers, support vector machines, decision trees, and neural networks (especially Graph Neural Networks, GNNs, that attracted a lot of attention recently). However, for classifiers to perform well, they must have access to a set of highly informative node features that can discriminate representatives of different classes. No matter how sophisticated classifiers one builds, they will perform poorly if they do not get informative input concerning the problem. In particular, although GNNs can aggregate features using network structure, they can benefit from good node-level features as an input~\cite{faber2021graph}. This is especially true for the features that cannot be computed via aggregation across neighbouring nodes. Hence, it is desirable to enrich a family of available features and apply machine learning tools to features of various sorts. 

Predictive models applied on high-dimensional data tend to overfit, which may cause performance degradation on unseen data (this issue is known as the curse of dimensionality)~\cite{hastie2009elements}. This problem can be solved with model regularization or various standard dimensionality reduction tools that can be categorized into two families: feature selection (selecting a subset of relevant features for model construction) and feature extraction (projecting the original high-dimensional features to a new space with low dimensionality)~\cite{li2017feature,tang2014feature}. Hence, the more node features potentially encapsulating additional, jointly non-redundant information, the better, as there exist efficient fitting procedures that can efficiently handle large sets of potential model features.

\medskip

In this paper, we investigate a family of features that pay attention to community structure often present in social networks~\cite{alhajj2018}. Community structure plays an important role in social network formation, and thus, it can be associated with nodes' properties. Such features are further called \emph{community-aware features}. Indeed, the community structure of real-world networks often reveals the internal organization of nodes~\cite{fortunato2010}. In social networks, communities may represent groups by interest; in citation networks, they correspond to related papers; in the Web, communities are formed by pages on related topics, etc. Such communities form groups of densely connected nodes with substantially fewer edges touching other parts of the graph. Identifying communities in a network can be done unsupervised and is often the analysts' first step. A better understanding of the network's community structure can be leveraged in its later analysis.

The motivation to study community-aware features is twofold. On the one hand, one can expect that such features can be highly informative for many node classification tasks. For example, it might be important whether a given node is a strong community member or, conversely, it is loosely tied to many communities. In terms of classical statistics, one can think of it as a question of whether a vector in a real space (representing one observation) belongs in a ``dense'' (strong member of some cluster) or ``sparse'' (data point between some clusters) region. Such problems are often studied by data depth methods~\cite{dd2002}. On the other hand, one can expect that community-aware features are not highly correlated to other features typically computed for networks. Indeed, to compute community-aware features, one needs first to identify the community structure of a graph. This, in turn, is a complicated non-linear transformation of the input graph, which cannot be expected to be easily recovered by supervised or unsupervised machine learning models that are not designed to be community-aware.

The contributions of this paper are the following:
\begin{itemize}
    \item We propose a new set of community-aware node features.
    \item We verify that the information in the proposed features is non-redundant against classical node features and against node embeddings (both classical and structural).
    \item We verify that the proposed features have predictive power in node classification tasks.
\end{itemize}

We show that there are classes of node prediction problems in which community-aware features have high predictive power. We also verify that community-aware features contain information that cannot be recovered either by classical node features or node embeddings (both classical and structural). In our experiments, we concentrate on binary classification to ensure that the results can be reported consistently across different graphs. We test our approach both on synthetic and real-life graphs. However, the same qualitative conclusions regarding the usefulness of community-aware features hold for problems involving multi-class or continuous target prediction.

\medskip

There are some community-aware features already introduced in the literature, such as CADA~\cite{Helling2018ACA} or the participation coefficient~\cite{guimera2005functional}; see Section~\ref{sec:varying} for their definitions. CADA is a feature that was originally developed as a measure of outlingness with respect to the community structure. On the other hand, the participation score measures how the neighbours of a given node are spread among communities.

However, it is important to highlight that both CADA and the participation score ignore the distribution of community sizes. We argue that considering community sizes when computing community-aware features matters as it provides a more detailed picture. As an example, consider a graph that has two communities; one of them contains 80\% of the total volume and the other 20\%. Now, consider two nodes, the first one has 80\% of its neighbours in the first community, and the second one has 80\% of neighbours in the second one. Under CADA and the participation score, they will have the same value of these metrics (as they ignore community sizes). However, we postulate that the first node is qualitatively different than the second one. Indeed, since the first community randomly has 80\% of the volume (e.g., under the Chung-Lu or the configuration models), one would expect both to have 80\% neighbours in the first community. The first node behaves exactly as expected. On the other hand, the behaviour of the second node is surprising. Most of its neighbours belong to small communities. Therefore, in this paper, we propose a class of community-aware features that, via the appropriate null model, consider community sizes and compare their predictive performance to the measures previously proposed in the literature.

\medskip

The paper is an extended version of the proceedings paper~\cite{short_version}\footnote{The proceedings paper~\cite{short_version} is significantly shorter. In particular, it does not contain Sections 2, 4.3, 4.4.2, and 5 from this paper, it does not provide derivation and detailed discussion of $\beta^*$ introduced in Section~3.2, and in Section~4, it does not contain the results presented for the \textbf{ABCD+o} graphs. We also present additional results for a larger real-life \textbf{Twitch} graph, which is not covered in the proceedings version.} and is structured as follows. In Section~\ref{sec:community}, we introduce the concept of null models that we will use to benchmark how strongly given node is attached to its community. In particular, we show how it is used to define modularity function, a quality function used by many clustering algorithms (Subsection~\ref{sec:modularity_function}). In Section~\ref{sec:varying}, we recall a few community-aware node features, CADA (Subsection~\ref{sec:CADA}), normalized within-module degree and participation coefficient (Subsection~\ref{sec:participation_score}), before introducing our own features, community association strength (Subsection~\ref{sec:beta_star}) and distribution-based measures (Subsection~\ref{sec:distribution_based}). Experiments are presented in Section~\ref{sec:experiments}. Three types of experiments were performed and reported: 
information overlap between community-aware and classical features (Subsection~\ref{sec:experiment1}),
one-way predictive power of community-aware and classical features (Subsection~\ref{sec:experiment2}),
and combined variable importance for prediction of community-aware and classical features (Subsection~\ref{sec:experiment3}).

\section{Using Null Models to Understand Community Structure}\label{sec:community}

A null model is a type of a random object that matches a specific property $\mathcal{P}$ observed is some dataset (for example, a collection of constraints such as a degree distribution in a graph following a given sequence $(d_i)_{i=1}^n$), but is otherwise taken randomly and unbiasedly from some larger family of objects having property $\mathcal{P}$ (in our previous example, all graphs of the same order and the degree distribution $(d_i)_{i=1}^n$). Using null models as a reference is a flexible approach for statistically testing the presence of properties of interest in empirical data.  As a result, the null models can be used to test whether a given object exhibits some ``surprising'' property that is not expected based on chance alone or as an implication of the fact that the object has property $\mathcal{P}$. A classical application of null models is frequentist hypothesis testing in statistics, where null-models are used to derive the distribution of statistics of interest under null hypothesis.

Null models were successfully used in network science to build various machine learning tools such as clustering algorithms~\cite{blondel2008fast,traag2019louvain} or unsupervised frameworks to evaluate node embeddings~\cite{kaminski2022multi,kaminski2020unsupervised}; see~\cite{matwin2023survey} for more examples.

\medskip

In this paper, we consider null models in the context of graphs. Null models play a central role not only in extracting graph community structure but, more importantly, they are used to quantify how tightly nodes are connected to the communities that surround them. More specifically, we consider null models that have the property $\mathcal{P}$ that ensures the degree distribution follows a given sequence observed in an empirical graph that we aim to analyze (either exactly, as in the configuration model~\cite{bollobas1980probabilistic,wormald1999models}, or in expectation, as in the Chung-Lu model~\cite{chung2006complex}). Insisting on property $\mathcal{P}$ is needed to make sure high degree nodes induce more edges than low degree ones under the null model. On the other hand, under the null model, there are no built-in communities, so edges are wired randomly as long as the degree distribution is preserved. As a result, such null models can be successfully used to benchmark and formally quantify how ``surprising'' it is to see that there are communities present in an empirical graph and that nodes are strongly attached to them.

To illustrate how null models are applied, in Subsection~\ref{sec:modularity_function} we define the modularity function that is a key ingredient in Leiden~\cite{traag2019louvain}, the clustering algorithm we use to extract community structure. However, the community-aware features we propose in Section~\ref{sec:varying} work also for other methods of community detection as long as they return a partition of nodes into communities. Clearly, they also work in the cases when ground-truth partitions are additionally provided.

\subsection{Modularity Function}\label{sec:modularity_function}

Consider a simple, unweighted graph $G=(V,E)$, where $V$ is a set of nodes and $E$ is a set of edges between nodes. Each edge $e\in E$ is a two-element subset of $V$. Given a subset of nodes $A\subseteq V$, we define the number of edges in the graph induced by this set (that is, the number of edges in $G$ that have both endpoints in $A$) as $e(A) = |\{a\in E: a\subseteq A\}|$. In particular, we have $e(V) = |E|$.

For any node $v \in V$, we define its degree as the number of neighbours of $v$ (that is, the number of edges that contain $v$): $\deg(v) = |\{e\in E: v \in e\}|$. For any subset of nodes $A\subseteq V$, we define its volume as the sum of degrees of nodes in $A$, that is, $\vol(A) = \sum_{v\in A}\deg(v)$. In particular, $\vol(V) = 2 |E|$. Finally, we say that $\A = \{A_1, A_2, \ldots, A_{\ell}\}$ is a partition of $V$ into $\ell$ sets if $A_i\cap A_j=\emptyset$ for any $1 \le i < j \le \ell$ and  $\bigcup_{i\in[\ell]} A_i = V$, where $[\ell] = \{1, 2, \ldots, \ell\}$.

With these definitions at hand, we are ready to define the modularity function of any partition $\A$ of $V$. The standard \emph{modularity function}, first introduced by Newman and Girvan in~\cite{newman2004finding}, is defined as follows:
\begin{equation}\label{eq:modularity}
    q(\A) = \sum_{A_i \in \A} \frac{e(A_i)}{|E|} - \sum_{A_i \in \A} \left( \frac{\vol(A_i)}{\vol(V)} \right)^2.
\end{equation}
The first term in~(\ref{eq:modularity}), $|E|^{-1} \sum_{A_i \in \A} e(A_i)$, is called the \emph{edge contribution} and it computes the fraction of edges that are captured within communities in partition $\A$. The second term in~(\ref{eq:modularity}), namely $\vol(V)^{-2} \sum_{A_i \in \A} \vol(A_i)^2$, is called the \emph{degree tax} and it computes the expected fraction of edges that do the same in the corresponding Chung-Lu~\cite{chung2006complex} null-model. In this random graph, the probability that node $v$ is adjacent to node $w$ (with loops allowed) is equal to $p(v,w) = \deg(v) \deg(w) / \vol(V)$ so that the expected degree of $v$ is equal to $\sum_{w \in V} p(v,w) = \deg(v)$, as desired. The modularity measures the deviation between the two.

The maximum modularity $q^*(G)$ is defined as the maximum of $q(\A)$ over all possible partitions $\A$ of $V$. It is used as a quality function by many popular clustering algorithms such as Louvain~\cite{blondel2008fast} and Leiden~\cite{traag2019louvain} that perform very well. It also provides an easy way to measure the presence of community structure in a network. If $q^*(G)$ is close to 1 (which is the trivial upper bound), we observe a strong community structure; conversely, if $q^*(G)$ is close to zero (which is the trivial lower bound, since $q(\A) = 0$ if $\A = \{V\}$), there is no community structure. 

\medskip

As already discussed in the introduction, the modularity function is prone to the so-called resolution limit reported in~\cite{fortunato2007resolution}. This means that an optimal partition of a large graph cannot contain small communities, that is, all $|A_i|$ are large. To overcome this problem, a simple modification of the modularity function was proposed (see e.g.~\cite{Lancichinetti2011}) that introduces the resolution parameter $\lambda>0$:
\begin{equation}\label{eq:modularity_lambda}
    q_{\lambda}(\A) = \sum_{A_i \in \A} \frac{e(A_i)}{|E|}  - \lambda\sum_{A_i \in \A} \left( \frac{\vol(A_i)}{\vol(V)} \right)^2.
\end{equation}
In this variant, if $\lambda$ is set to be larger than $1$, then large communities (communities for which $\vol(A_i)$ is large) are penalized more than small ones. As a result, partitions $\A$ that yield large $q_{\lambda}(\A)$ tend to consist of increasingly smaller communities as $\lambda$ grows. If $\lambda\to\infty$, then the edge contribution of $q_{\lambda}(\A)$ becomes negligible and the optimal partition turns out to be $\A = \{\{v\}: v\in V\}$ in which each node creates its own single node community.

For more details and discussion of alternative approaches, we direct the reader to~\cite{Reichardt2006,xiang2012} or any book on complex networks such as~\cite{kaminski2021mining,lambiotte2021modularity}.

\section{Community-Aware Node Features}\label{sec:varying}

In this section, we introduce various community-aware node features. All of them aim to capture and quantify how given nodes are attached to communities. It will be assumed that a partition $\A = \{A_1, A_2, \ldots, A_{\ell}\}$ of $V$ into $\ell$ communities is already provided; communities induced by parts $A_i$ ($i \in [\ell]$) are denser comparing to the global density of the graph. Such partition can be found by any clustering algorithm. In our empirical experiments we use Leiden~\cite{traag2019louvain} which is known to produce good, stable results.

To simplify the notation, we will use $\deg_{A_i}(v)$ to be the number of neighbours of $v$ in $A_i$, that is, $\deg_{A_i}(v) =  | N(v) \cap A_i |$, where $N(v)$ is the set of neighbours of $v$. 

\medskip

We start with three node features that have been already proposed in the literature: anomaly score CADA (Subsection~\ref{sec:CADA}), normalized within-module degree and participation coefficient that usually work in tandem (Subsection~\ref{sec:participation_score}). As mentioned in the introduction, these three features completely ignore community sizes in their definitions. Using Chung-Lu model~\cite{chung2006complex} as the null model, we can easily incorporate this useful information. We propose a few new community-aware features that take this into account: community association strength (Subsection~\ref{sec:beta_star}) and various distribution-based measures (Subsection~\ref{sec:distribution_based}).

\subsection{Anomaly Score CADA}\label{sec:CADA}

The first community-aware node feature is the anomaly score introduced in~\cite{helling2019community} with the goal to describe to what extent the neighbours of a node belong to a diverse number of communities, while the node itself does not strongly belong to one of them. The \emph{anomaly score} is computed as follows: for any node $v \in V$ with $\deg(v) \ge 1$,
\begin{equation*}
\cd(v) = \frac {\deg(v)} {d_{\A}(v)}, \qquad \text{ where } \qquad d_{\A}(v) = \max \Big\{ \deg_{A_i}(v) : A_i \in \A \Big\};
\end{equation*}
the denominator, $d_{\A}(v)$, represents the maximum number of neighbouring nodes that belong to the same community. In one extreme, if all neighbours of $v$ belong to the same community, then $\cd(v) = 1$. In the other extreme, if no two neighbours of $v$ belong to the same community, then $\cd(v) = \deg(v)$.

Note that $\cd(v)$ does not pay attention to which community node $v$ belongs to. Moreover, this node feature is unbounded, that is, $\cd(v)$ may get arbitrarily large. As a result, we will also investigate the following small modification of the original score, the \emph{normalized anomaly score}: for any node $v \in A_i$ with $\deg(v) \ge 1$,
\begin{equation*}
\overline{\cd}(v) = \frac { \deg_{A_i}(v) }{\deg(v)}. 
\end{equation*}
Clearly, $0 \le \overline{\cd}(v) \le 1$. Moreover, any good clustering algorithm typically should try to assign $v$ to the community where most of its neighbours are, so most nodes are expected to have $\overline{\cd}(v) = 1 / \cd(v)$. The case when this condition might not hold is if some node has slightly more neighbours in some large community than in some small community. Indeed, it might happen that a community detection algorithm maximizing the modularity function assigns some node to a small community despite the fact that it is not a community where the node has most of its neighbours in. For example, consider the situation in which there are two communities of respective sizes 80\% and 20\% of the total volume, and a node that has 51\% of its neighbours in large community and 49\% of its neighbours in a small community. This also shows the importance of paying attention to community sizes.

\subsection{Normalized Within-module Degree and Participation Coefficient}\label{sec:participation_score}

In~\cite{guimera2005functional}, an interesting and powerful approach was proposed to quantify the role played by each node within a network that exhibits community structure. Seven different universal roles were heuristically identified, each defined by a different region in the $(z(v), p(v))$ 2-dimensional parameter space, where $z(v)$ is the normalized within-module degree of a node $v$ and $p(v)$ is the participation coefficient of $v$. Node feature $z(v)$ captures how strongly a particular node is connected to other nodes within its own community, completely ignoring edges between communities. On the other hand, node feature $p(v)$ captures how neighbours of $v$ are distributed between all parts of the partition $\A$.

Formally, the \emph{normalized within-module degree} of a node $v$ is defined as follows: for any node $v \in A_i$,
$$
z(v) = \frac {\deg_{A_i}(v) - \mu(v)}{\sigma(v)},
$$
where $\mu(v)$ and $\sigma(v)$ are, respectively, the mean and the standard deviation of $\deg_{A_i}(u)$ over all nodes $u$ in the community $v$ belongs to. Note that in the definition above we assumed that the graph induced by the community node $v$ belongs to is \emph{not} regular (that is, $\sigma(v) \neq 0$). In our numerical experiments, if $\sigma(v)=0$, then we simply take $z(v)=0$. This situation might happen in practice when a small community is detected, since it is highly unlikely for a large set of nodes to induce a regular graph. Note also that $z(v)$ is the familiar $Z$-score as it measures how many standard deviations the internal degree of $v$ deviates from the mean. If node $v$ is tightly connected to other nodes within the community, then $z(v)$ is large and positive. On the other hand, $|z(v)|$ is large and $z(v)$ is negative when $v$ is loosely connected to other peers. 

The \emph{participation coefficient} of a node $v$ is defined as follows: for any node $v \in V$ with $\deg(v) \ge 1$,
$$
p(v) = 1 - \sum_{i = 1}^{\ell} \left( \frac {\deg_{A_i}(v)}{\deg(v)} \right)^2.
$$
The participation coefficient $p(v)$ is equal to zero if $v$ has neighbours exclusively in some community (most likely in its own community). In the other extreme situation, the neighbours of $v$ are homogeneously distributed among all parts and so $p(v)$ is close to the trivial upper bound of
$$
1 - \sum_{i = 1}^{\ell} \left( \frac {\deg(v)/\ell}{\deg(v)} \right)^2 = 1 - \frac {1}{\ell} \approx 1
$$
which is close to 1 for large $\ell$.

\subsection{Community Association Strength}\label{sec:beta_star}

As already advertised, let us now introduce our own community-aware node feature that takes the distribution of community sizes into account. In order to build an intuition, suppose for a moment that we aim to adjust the modified modularity function~(\ref{eq:modularity_lambda}) to detect nodes that are outliers. If the fraction of neighbours of a node $v$ in its own community is small relative to the corresponding expected fraction under the null-model, then we will say that $v$ is likely to be an outlier. In other words, in order to quantify the probability that $v$ is an outlier, one might want to compare $\deg_{A_i}(v)/\deg(v)$ against $\vol(A_i)/\vol(V)$. Our goal is to adjust the modularity function in such a way that nodes that are likely to be outliers are put into single node communities.

Let us formalize these concepts. Given a partition $\A$, we define a set of outliers as $O=\bigcup_{i \in [\ell] : |A_i|=1} A_i$, that is, nodes that are put into a single node communities are defined as outliers. For a fixed parameter $\beta\geq 0$ (and the resolution parameter $\lambda>0$), we define the regularized modularity function as follows:
\begin{eqnarray}\label{eq:modularity_beta}
q_{\lambda,\beta}(\A) &=& \sum_{A_i \in \A} \frac{e(A_i) + \delta_{|A_i|=1} \beta \, \vol(A_i)/2}{|E| + Z/2} -
\lambda\left(\sum_{A_i \in \A} \left( \frac{\vol(A_i)(1+\delta_{|A_i|=1} \beta)}{\vol(V) + Z} \right)^2\right),
\end{eqnarray}
where $Z=\sum_{A_i \in \A} \delta_{|A_i|=1} \beta \, \vol(A_i) = \beta \, \vol(O)$; $\delta_B$ is the Kronecker delta: $\delta_B=1$ if $B$ is true and  $\delta_B=0$, otherwise.

The above definition clearly generalizes the modified modularity function~(\ref{eq:modularity_lambda}), and we recover it when $\beta = 0$. For $\beta > 0$, the rationale behind it is as follows. 
If additional self-loops are introduced in graph $G$ in communities containing outliers (single node communities), then the number of them is guided by parameter $\beta$ and is proportional to the volume of such small communities (but not their node count, as they contain only one node).
This impacts the edge contribution and the degree tax is adjusted accordingly. If $\beta$ is close to 0, then only nodes that are loosely attached to their own communities are pushed to single communities (and so they become outliers) since such operation increases the modularity function~(\ref{eq:modularity_beta}). The larger $\beta$ gets, the more nodes have incentive to become outliers. Similarly to the original modularity function, particular values of $q_{\lambda,\beta}(\A)$ are not interpretable. It is designed for algorithms such as Louvain or Leiden, trying to maximize function $q_{\lambda,\beta}(\A)$, to find outliers. In our application it will be used to define node features.

Unfortunately, the formula~(\ref{eq:modularity_beta}) is challenging to work with, since the modifications are affecting the numerators and the denominators of both the edge contribution and the degree tax. It is easier to use the following approximation instead: 
\begin{equation*}
q_{\lambda,\beta}(\A) \approx \sum_{A_i \in \A} \frac{e(A_i) + \delta_{|A_i|=1} \beta\vol(A_i)/2}{|E|} -
\lambda\left(\sum_{A_i \in \A} \left( \frac{\vol(A_i)}{\vol(V)} \right)^2\right).
\end{equation*}

Using this approximation, one can ask the following question for any node $v$ that belongs to community $A_i$: what is the threshold value of $\beta^*(v)$ so that if $\beta > \beta^*(v)$, then the approximation of the regularized modularity function increases if $v$ is moved from $A_i$ to form its own, single node community. The approximated version can be easily analyzed to get such threshold. Indeed, the change associated with the edge contribution $\sum_{A_i \in \A} {(e(A_i) + \delta_{|A_i|=1} \beta\vol(A_i)/2)}/{|E|}$ when we remove node $v$ from its community $A_i$ and put it into a new community that contains only this node is equal to
\begin{equation}\label{eq:beta_edge_contr}
\frac {-\deg_{A_i}(v) + \beta \deg(v)/2}{|E|} = \frac {-2\deg_{A_i}(v) + \beta \deg(v)}{ \vol(V) }
\end{equation}
whereas the change associated with the degree tax $-\lambda\left(\sum_{A_i \in \A} \left({\vol(A_i)}/{\vol(V)} \right)^2\right)$ when we remove node $v$ from its community $A_i$ and put it into a new community that contains only this node is
\begin{equation}\label{eq:beta_degree_tax}
\lambda \frac {( \vol(A_i) - \deg(v) )^2 + \deg(v)^2 - \vol(A_i)^2 }{ \vol(V)^2 } = -2 \lambda \frac { \vol(A_i) \deg(v) - \deg(v)^2 }{ \vol(V)^2 } \,.
\end{equation}
The threshold value may be then computed by finding the unique value of $\beta$ that makes~(\ref{eq:beta_edge_contr}) equal to~(\ref{eq:beta_degree_tax}). Hence, for any $v \in A_i$, we define the \emph{community association strength} as follows:
$$
\beta^*(v)=2 \left( \frac{\deg_{A_i}(v)}{\deg(v)} - \lambda\frac{\vol(A_i)-\deg(v)}{\vol(V)} \right).
$$
The lower the value of $\beta^*(v)$, the less associated node $v$ with its own community is. In the derivation above we allow for any $\lambda > 0$, but in the experiments, we will use $\lambda=1$.

Let us also notice that when $\lambda=1$, $\beta^*(v)$ is essentially twice the normalized anomaly score $\overline{\cd}(v)$ after adjusting it to take into account the corresponding prediction from the null model. Moreover, let us note that some simplified version of this node feature was already used in~\cite{kaminski2023artificial}.

\medskip

\begin{figure}[h]
\centering
\includegraphics[width=0.6\textwidth]{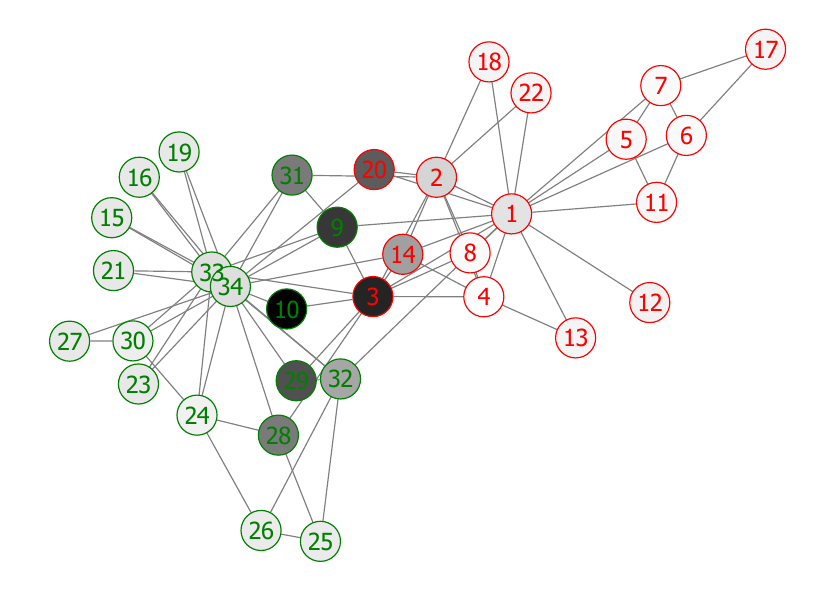}
\caption{Communities (red and green colours) in the Karate graph. The shades of nodes correspond to their values of $\beta^*(v)$ (darker colours indicate lower values).}\label{fig:karate}
\end{figure}

To illustrate the usefulness of this new node feature on a toy example, we consider the well-known Karate Club graph~\cite{karate} in Figure~\ref{fig:karate}. There are two ground truth communities which can be distinguished by red and green node colours. The shades of nodes correspond to the values of the community association strength $\beta^*(v)$; darker shades indicate lower values of this node feature. We see that nodes $3$ and $10$ are the darkest and, indeed, they have the same number of neighbours in their own community as outside of it. Also, in general, we see that darker nodes are in the middle of the plot, at the ``intersection'' of communities, while light nodes are on the left and right borders (they have all neighbours within their own communities). It is important to notice that to layout the graph we used a standard force-directed algorithm that assumes that some kind of attractive forces (imagine springs connecting nodes) are used to attract nodes connected by edges together, while simultaneously repulsive forces (imagine electrically charged particles) are used to separate the remaining pairs of nodes. As a result, ``tightly'' connected clusters of nodes will show up close to each other, and those that are ``loosely'' connected will be repulsed towards the outside. The fact that $\beta^*(v)$ was able to recover the position of nodes is a good and promising sign. Other community-aware features should produce similar results for this graph as its two ground truth communities have similar sizes. 

\subsection{Distribution-Based Measures}\label{sec:distribution_based}

Our next community-aware node features are similar in spirit to the participation coefficient, that is, they aim to measure how neighbours of a node $v$ are distributed between all parts of the partition $\A$. The main difference is that they pay attention to the sizes of parts of $\A$ and compare the distribution of neighbours to the corresponding predictions from the null model. They are upgraded versions of the participation coefficient, similarly to the community association strength being an upgraded counterpart of the normalized anomaly score.

Formally, for any node $v \in V$, let $q_1(v)$ be the vector representing fractions of neighbours of $v$ in various parts of partition $\A$, that is, 
$$
q_1(v) = \left( \frac {\deg_{A_1}(v)}{\deg(v)}, \frac {\deg_{A_2}(v)}{\deg(v)}, \ldots, \frac {\deg_{A_{\ell}}(v)}{\deg(v)} \right).
$$
Similarly, let $\hat{q}_1(v)$ be the corresponding prediction for the same vector based on the Chung-Lu model, that is,
$$
\hat{q}_1(v) = \left( \frac {\vol(A_1)}{\vol(V)}, \frac {\vol(A_2)}{\vol(V)}, \ldots, \frac {\vol(A_{\ell})}{\vol(V)} \right) =: \hat{q}_1.
$$
Note that $\hat{q}_1(v) = \hat{q}_1$ does \emph{not} depend on $v$ (of course, it should not!) but only on the distribution of community volumes. Our goal is to measure how similar the two vectors are. A natural choice would be any of the $p$-norms but, since both vectors are stochastic (that is, all entries are non-negative and they add up to one), alternatively one can also use any good measure for comparison of probability distributions. In our experiments we tested the following node features:
\begin{itemize}
    \item \emph{$L^1$ norm}: $L^1_1(v) = \sum_{i=1}^{\ell} \left| \frac {\deg_{A_i}(v)}{\deg(v)} - \frac {\vol(A_i)}{\vol(V)} \right|$
    \item \emph{$L^2$ norm}: $L^2_1(v) = \left( \sum_{i=1}^{\ell} \left( \frac {\deg_{A_i}(v)}{\deg(v)} - \frac {\vol(A_i)}{\vol(V)} \right)^2 \right)^{1/2}$ 
    \item \emph{Kullback--Leibler divergence}~\cite{csiszar1975divergence}: $\kl_1(v) = \sum_{i=1}^{\ell} \frac {\deg_{A_i}(v)}{\deg(v)} \log \left( \frac {\deg_{A_i}(v)}{\deg(v)} \cdot \frac {\vol(V)}{\vol(A_i)} \right)$
    \item \emph{Hellinger distance}~\cite{hellinger1909neue}: $h_1(v) = \frac {1}{\sqrt{2}} \left( \sum_{i=1}^{\ell} \left( \left( \frac {\deg_{A_i}(v)}{\deg(v)} \right)^{1/2} - \left( \frac {\vol(A_i)}{\vol(V)} \right)^{1/2} \right)^2 \right)^{1/2}$ 
\end{itemize}

The above measures pay attention to which communities neighbours of $v$ belong to. However, some of such neighbours might be strong members of their own communities but some of them might not be. Should we pay attention that? Is having a few strong members of community $A_i$ as neighbours equivalent to having many neighbours that are weak members of $A_i$? To capture these nuances, one needs to consider larger ego-nets around $v$, nodes at distance at most 2 from $v$. We define $q_2(v)$ to be the average value of $q_1(u)$ taken over all neighbours of $v$, that is,
$$
q_2(v) = \frac {1}{\deg(v)} \sum_{u \in N(v)} q_1(u).
$$
As before, $\hat{q}_2(v)$ is the corresponding prediction based on the null model. However, since $\hat{q}_1(u) = \hat{q}_1$ does not depend on $u$, $\hat{q}_2(v)$ also does not depend on $v$ and, in fact, it is equal to $\hat{q}_1$. The difference between $q_2(v)$ and $\hat{q}_2(v)$ may be measured by any metric used before. In our experiments we tested $L^1_2(v)$, $L^2_2(v)$, $\kl_2(v)$, and $h_2(v)$, counterparts of $L^1_1(v)$, $L^2_1(v)$, $\kl_1(v)$, and $h_1(v)$ respectively. 

Let us mention that $q_1(v)$ and $q_2(v)$ have a natural and useful interpretation. Consider a random walk that starts at a given node $v$. The $i$th entry of the $q_1(v)$ vector is the probability that a random walk visits a node from community $A_i$ after one step. Vector $q_2(v)$ has the same interpretation but after two steps are taken by the random walk. 

One can repeat the same argument and define $L^1_i(v)$, etc., for any natural number $i$ by performing $i$ steps of a random walk. Moreover, a natural alternative approach would be to consider all possible walk lengths but connections made with distant neighbours are penalized by an attenuation factor $\alpha$ as it is done in the classical Katz centrality~\cite{katz1953new}. 

Finally, let us note that the above aggregation processes could be viewed as simplified versions of GNNs classifiers. Therefore, the investigation of these measures additionally shows how useful community-aware measures could be when used in combination with GNN models.

\section{Experiments}\label{sec:experiments}

\subsection{Graphs Used}

We consider undirected, connected, and simple (no loops nor parallel edges are allowed) graphs so that all node features are well defined and all methods that we use work properly. In each graph, we have some ``ground-truth'' labels for the nodes which is used to benchmark classification algorithms. For consistency of the reported metrics, we consider binary classification tasks, so the ground-truth node features that are to be predicted will always consist of labels from the set $\{0, 1\}$ with label~$1$ being the target class. 
We consider generic binary classification, and the choice of classes will vary for different experiments.

In the experiments, we used two families of graphs. The first family consists of synthetic networks. 
The main goal of experiments on this family is to show the added value of community-aware node features.
In these networks, the target class depends on the overall community structure of the graph. \textbf{A}rtificial \textbf{B}enchmark for \textbf{C}ommunity \textbf{D}etection with \textbf{O}utliers (\textbf{ABCD+o})~\cite{kaminski2023artificial} fits this need perfectly. Nodes in these synthetic graphs have binary labels: community-aware outliers (with label $1$) do not belong strongly to any of the communities whereas other nodes (with label $0$) are members of a community, and we can control the strength of such memberships.

The second family of networks we used in our experiments are empirical real-world graphs (mainly social networks, but also other types of networks for completeness of the analysis). We tried to select a collection of graphs with different properties (density, community structure, degree distribution, clustering coefficient, etc.). More importantly, some of them have highly unbalanced binary classes. Experiments with these networks will serve as a more challenging and robust test for usefulness of the proposed community-aware node features.

\subsubsection{Synthetic ABCD+o Graphs}\label{sec:abcd}

The \textbf{A}rtificial \textbf{B}enchmark for \textbf{C}ommunity \textbf{D}etection graph (\ABCD)~\cite{kaminski2021artificial} is a random graph model with community structure and power-law distribution for both degrees and community sizes. The model generates graphs with similar properties as the well-known \textbf{LFR} model~\cite{lancichinetti2009benchmarks,lancichinetti2008benchmark}, and its main parameter $\xi$ (counterpart of the mixing parameter $\mu$ in the \textbf{LFR} model) controls the level of noise, that is, the proportion of edges that touch two distinct communities. Both models produce synthetic networks with comparable properties but \textbf{ABCD} is significantly faster than \textbf{LFR} (especially its fast implementation that uses multiple threads, \textbf{ABCDe}~\cite{kaminski2022properties}) and can be easily tuned to allow the user to make a smooth transition between the two extremes: pure (disjoint) communities and random graph with no community structure. Moreover, it is easier to analyze theoretically. For example, various theoretical asymptotic properties of the \ABCD\ model are analyzed in~\cite{kaminski2022modularity}, including the modularity function that is an important graph property of networks in the context of community detection.

An important feature of the family of \ABCD\ networks is its flexibility. Hypergraph counterpart of the model, \hABCD~\cite{kaminski2022hypergraph}, was recently introduced that can mimic any desired level of homogeneity of hyperedges that fall into one community. More importantly from the perspective of the current paper, an extension of the \ABCD\ model to include community-aware outliers, \textbf{ABCD+o}, was introduced in~\cite{kaminski2023artificial}. The outlier nodes in this model are not assigned to any community; their neighbours are sampled from the entire graph. Experiments in~\cite{kaminski2023artificial} were performed to show that outliers in the new model as well as outliers in real-world networks pose similar distinguishable properties which ensures that it may potentially serve as a benchmark of outlier detection algorithms.

In our experiments, we generated \textbf{ABCD+o} networks on $n=10{,}000$ nodes, including $s_0=1{,}000$ outliers (10\%).
The degree distribution follows a power-law with exponent $\gamma = 2.5$ and degrees are between 5 and 500. The distribution of community sizes follows a power-law with exponent $\beta = 1.5$ and their sizes range from $50$ to $2{,}000$. We generated 4 networks with different level of noise: $\xi\in\{0.3, 0.4, 0.5, 0.6\}$. The lower the value of $\xi$, the more tight the communities are which makes it easier to detect communities as well as to identify outliers.


\subsubsection{Empirical Graphs}\label{sec:real}

For experiments on real-world, empirical networks, we selected the following six datasets. In the selection process we focused on social networks (four data sets), but also, for completeness of the analysis, included two networks of other types. 
One of the graphs (\textbf{Twitch}) is larger than the others so we can additionally test the scalability of the proposed methods. In cases when multiple connected components were present, we kept only the giant component. Self-loops, if present, were also dropped before performing the experiments.
We summarize some statistics for the above graphs in Table~\ref{tab:real_networks}.
The number of communities reported in the table are communities identified by running the Leiden algorithm $1{,}000$ times independently on a respective graph and picking the community partition with the highest modularity (see Section~\ref{sec:all_features} for more details).

\begin{itemize}

\item \textbf{Reddit}~\cite{kumar2019predicting}: A user-subreddit graph which consists of one month of posts made by users on subreddits. This is a bipartite graph with $9{,}998$ nodes representing users in one part and $982$ nodes representing subreddits in the other one. This dataset contains ground-truth labels of banned users from Reddit which we use as the target class (label~1). Nodes associated with subreddits are not used for training nor evaluation but are kept for building node features associated with users. 

\item \textbf{Grid}~\cite{SciGRIDv0.2}:  A European high-voltage power grid, extracted in 2016 by GridKit from OpenStreetMap. Nodes correspond to stations and edges represent lines between stations. Nodes in the original data set have attributes such as ``joint'', ``merge'', ``plant'', ``station'' and ``substation''. For the target class we selected the attribute ``plant'' because it was the least frequent attribute in the data, so we can have a test in which the target is highly unbalanced.

\item \textbf{Facebook}~\cite{rozemberczki2021multi}: In this graph, nodes represent official Facebook pages while the edges are mutual likes between sites. Nodes are labelled by Facebook and belong to one of the 4 categories: politicians, governmental organizations, television shows and companies; we selected politicians as our target class. 

\item \textbf{LastFM}~\cite{feather}: A social network of LastFM users from Asian countries. Nodes are associated with users and edges are mutual follower relationships between them. The node features were extracted based on the artists liked by the users. The network was designed with multinomial node classification in mind: one has to predict the location of users. For our purpose, we ignore all node features but the country field and use ``country 17'' (the most frequent country because we wanted to have a test in which the target would not be highly unbalanced) as the target class. 

\item \textbf{Amazon}~\cite{dou2020enhancing}: This dataset includes product reviews on Amazon under the ``musical instruments'' category. Nodes in this graph are users and edges connect users that reviewed at least one common product. Users with with less than 20\% ``helpful'' votes are labelled as fraudulent entities (label 1) whereas users with at least 80\% helpful votes are labelled as benign entities (label~0). Some nodes have missing labels; as it was done in the case of Reddit network, we do not use them for training nor evaluation but we keep them for building node features of the labeled nodes. 

\item \textbf{Twitch}~\cite{rozemberczki2021twitch}: A social network of Twitch users which was collected from the public API in Spring 2018. 
Nodes are Twitch users and edges are mutual follower relationships between them. 
For this graph, the binary prediction task identifies if the user streams mature content (label~1) or gaming content (label~0).
\end{itemize}

\begin{table}[htp]
\caption{Statistics of the selected real-world empirical graphs.}
\begin{center}
\begin{tabular}{|c|c|c|c|c|c|}
\hline
dataset & \# & average & \#  & target & target \\
 & of nodes &  degree & of communities & proportion & description \\
\hline
\textbf{Reddit} & 10{,}980 & 14.30 & 12& 3.661\% & is node a banned user \\
\textbf{Grid} & 13{,}478 & 2.51 & 78 & 0.861\% & is node a plant \\
\textbf{LastFM} & 7{,}624 & 7.29 & 28 & 20.619\% & is node in country \#17 \\
\textbf{Facebook} & 22{,}470 & 15.20 & 58 & 25.670\% & is node a politician \\
\textbf{Amazon} & 9{,}314 & 37.49 & 39 & 8.601\% & is node fraudulent \\
\textbf{Twitch} & 168{,}114 & 80.87 & 19 & 47.01\% & is streamed content mature \\
\hline
\end{tabular}
\end{center}
\label{tab:real_networks}
\end{table}

\subsection{Node Features Investigated}\label{sec:all_features}

The community-aware node features that we tested are summarized in Table~\ref{tab:features1}. Their precise definitions can be found in Section~\ref{sec:varying}. The features are computed with reference to a partition of a graph into communities obtained using the Leiden algorithm. The partition is chosen as the best of $1{,}000$ independent runs of the \texttt{community\_leiden} function implemented in the \emph{igraph} library~\cite{igraph2006} (Python interface of the library was used). Each of such independent runs was performed until a stable iteration was reached.

\begin{table}[htp]
\caption{Community-aware node features used in our experiments. A combination of \texttt{WMD} and \texttt{CPC} is also used as a 2-dimensional embedding of a graph (\texttt{WMD+CPC}).}
\begin{center}
\begin{tabular}{|c|c|l|c|}
\hline
abbreviation & symbol & name & subsection \\
\hline
\texttt{CADA} & $\cd(v)$ & anomaly score CADA & \ref{sec:CADA} \\
\texttt{CADA*} & $\overline{\cd}(v)$ & normalized anomaly score & \ref{sec:CADA} \\
\texttt{WMD} & $z(v)$ & normalized within-module degree & \ref{sec:participation_score} \\
\texttt{CPC} & $p(v)$ & participation coefficient & \ref{sec:participation_score} \\
\texttt{CAS} & $\beta^*(v)$ & community association strength & \ref{sec:beta_star} \\
\texttt{CD\_L11} & $L^1_1(v)$ & $L^1$ norm for the 1st neighbourhood & \ref{sec:distribution_based} \\
\texttt{CD\_L21} & $L^2_1(v)$ & $L^2$ norm for the 1st neighbourhood & \ref{sec:distribution_based} \\
\texttt{CD\_KL1} & $\kl_1(v)$ & Kullback--Leibler divergence for the 1st neighbourhood & \ref{sec:distribution_based} \\
\texttt{CD\_HD1} & $h_1(v)$ & Hellinger distance for the 1st neighbourhood & \ref{sec:distribution_based} \\
\texttt{CD\_L12} & $L^1_2(v)$ & $L^1$ norm for the 2nd neighbourhood & \ref{sec:distribution_based} \\
\texttt{CD\_L22} & $L^2_2(v)$ & $L^2$ norm for the 2nd neighbourhood & \ref{sec:distribution_based} \\
\texttt{CD\_KL2} & $\kl_2(v)$ & Kullback--Leibler divergence for the 2nd neighbourhood & \ref{sec:distribution_based} \\
\texttt{CD\_HD2} & $h_2(v)$ & Hellinger distance for the 2nd neighbourhood & \ref{sec:distribution_based} \\
\hline
\end{tabular}
\end{center}
\label{tab:features1}
\end{table}

Classical (non-community-aware) node features are summarized in Table~\ref{tab:features2}. These are standard and well-known node features. We omit their precise definitions but, instead, refer to the appropriate sources in the table. Alternatively, their definitions can be found in any book on mining complex networks such as~\cite{kaminski2021mining}. 

Finally, we will use two more sophisticated and powerful node features obtained through graph embeddings, where a graph embedding is a mapping from a set of nodes of a graph into a real vector space.
Embeddings can have various aims like capturing the underlying graph topology and structure, node-to-node relationship, or other relevant information about the graph, its subgraphs or nodes themselves. Embeddings can be categorized into two main types: classical embeddings and structural embeddings. Classical embeddings focus on learning both local and global proximity of nodes, while structural embeddings learn information specifically about the local structure of nodes’ neighbourhood. We test one embedding from each class: \texttt{node2vec}~\cite{grover2016} and \texttt{struc2vec}~\cite{ribeiro2017}. The parameters used for the embeddings are as follows:
\begin{itemize}
    \item \texttt{node2vec}: dim=16; walk-length=50; num-walks=10; p=1; q=1.
    \item \texttt{struc2vec}: dim=16; num-walks=10; walk-length=50; window-size=5; OPT1, OPT2, and OPT3 set to true.
\end{itemize}

\begin{table}[htp]
\caption{Classical (non-community-aware) node features that are used in our experiments.}
\begin{center}
\begin{tabular}{|c|l|c|}
\hline
abbreviation & name & reference \\
\hline
\texttt{lcc} & local clustering coefficient & \cite{ws1998} \\
\texttt{bc} & betweenness centrality & \cite{freeman1977} \\
\texttt{cc} & closeness centrality & \cite{bavelas1950} \\
\texttt{dc} & degree centrality & \cite{kaminski2021mining} \\
\texttt{ndc} & average degree centrality of neighbours & \cite{barrat2004} \\
\texttt{ec} & eigenvector centrality &  \cite{bonacich2001} \\
\texttt{eccen} & node eccentricity & \cite{buckley1990} \\
\texttt{core} & node coreness & \cite{kaminski2021mining} \\
\texttt{n2v} & 16-dimensional \texttt{node2vec} embedding & \cite{grover2016} \\
\texttt{s2v} & 16-dimensional \texttt{struc2vec} embedding & \cite{ribeiro2017} \\
\hline
\end{tabular}
\end{center}
\label{tab:features2}
\end{table}

\subsection{Time complexity}

Given some (synthetic or empirical) graph under consideration, let $n$ be the number of nodes, $m$ the number of edges and $\ell$ the number of communities obtained with some algorithm. Recall that potential isolated nodes are removed before the experiments start and so we may assume that $m = \Omega(n)$. The major computational cost of computing the community-aware features comes from running the community detection algorithm. In our study we use the Leiden algorithm which, for sparse graphs, has an empirically verified $O(n\log n)$ running time for sparse graphs ($m=O(n)$).

Most community-aware measures defined above (except the second neighbourhood ones)
can be computed in $O(m+n\cdot \ell)$ time. First, we traverse all edges of the graph and aggregate the number of neighbours of all nodes in respective communities and next, using this information, we compute the desired measure which can be done in one pass over the data.

The second neighbourhood measures (that is, $L_2^1(v)$, $L_2^2(v)$, $\kl_2(v)$, and $h_2(v)$) require an additional step in each the averages over first neighbourhood measures of a given node are computed. This step can be done in $O(m\cdot \ell)$ time and thus the overall computational complexity is $O(m + n\cdot \ell + m \cdot \ell) = O(m\cdot \ell)$.

Most real-world networks are sparse ($m = O(n)$) and so in such networks almost all nodes have relatively small degrees (consider, for example, power-law networks which are quite common). As a result, efficient algorithms compute community-aware node features for such networks in linear time (one does not need to consider all $\ell$ communities for a single node but only communities a node is connected to; $O(1)$ of them, on average) which is much faster than the time required to run the Leiden algorithm. Moreover, computing community-aware features is often faster than classical ones. In particular, some classical features such as betweenness centrality have significantly worse complexity. Similarly, computation of \texttt{node2vec} and \texttt{struc2vec} embeddings is significantly more time consuming than computing community-aware features. This problem was especially visible for the largest graph considered (\textbf{Twitch}) for which the computation of these features took many hours, while computation of community-aware measures was fast.

\subsection{Results of the experiments}\label{sec:experiment_setup}

In this section, we present the results of three numerical experiments that were performed to investigate the usefulness of community-aware features:
\begin{enumerate}
    \item \emph{information overlap} between community-aware and classical features;
    \item \emph{one-way predictive power} of community-aware and classical features;
    \item \emph{combined variable importance for prediction} of community-aware and classical features.
\end{enumerate}
The details behind these experiments and the observations are provided in the independent subsections below. From the computational perspective, all analytical steps (generation of graphs, extractions of both community-aware and classical features, execution of experiments) were implemented in such a way that all experiments are fully reproducible. In particular, all steps that involve pseudo-random numbers were appropriately seeded. The source code allowing for reproduction of all results is available at GitHub repository\footnote{\url{https://github.com/sebkaz/BetaStar.git}}.

\subsubsection{Information Overlap}\label{sec:experiment1}

In the first experiment (\emph{information-overlap}), our goal was to test, using a variety of models, to what extent each community-aware feature described in Table~\ref{tab:features1} can be explained by all the classical features from Table~\ref{tab:features2} (including both embeddings, \texttt{node2vec} and \texttt{struc2vec}).

In this experiment each community-aware feature was a target in the model. The features were all classical features. Our goal was to check how well a given community-aware feature can be explained (predicted). As a measure of this prediction quality we used the Kendall correlation of the value of the target community-aware feature and its prediction produced by the model. We used the \texttt{kendaltau} function from the scipy python package\footnote{\url{https://docs.scipy.org/doc/scipy-1.12.0/reference/generated/scipy.stats.kendalltau.html}} which computes the Tau-b statistic that makes adjustments for ties in the input data. To ensure that the reported results are robust and capture possible non-linear relationships between combinations of classical features and a target community-aware feature, for each community-aware feature, five models were built using random forest, xgboost, lightgbm, linear regression, and regularized regression. The maximum Kendall correlation that was obtained is reported.

We used the non-parametric Kendall correlation to have a measure that is robust to possible non-linear relationships, since Kendall correlation checks how well the ordering of predictions matches the ordering of the target. Nevertheless, we also used the $R^2$ measure, which assumes linearity of the relationship. The results obtained were similar. The model building procedure assumed a random train-test split of nodes with a proportion of 70/30. The reported Kendall correlation values were computed on test data set.

The goal of this experiment is to show that community-aware features cannot be explained by classical features (including two highly expressible embeddings). The conclusion is that it is worth to include such features in predictive models as they could potentially improve their predictive power. 
However, this additional information could be simply a noise and so not useful in practice. To verify the usefulness of the community-aware features, we performed two more experiments, namely, \emph{one-way predictive power} and \emph{combined variable importance for prediction} checks. 
In these experiments, we check if community-aware features are indeed useful in node label prediction problems. 

In general, the expectation is that for synthetic networks such as \textbf{ABCD+o} graphs, the community-aware features should significantly outperform classical features. Indeed, recall from Section~\ref{sec:abcd} that the target variable in these networks is whether or not some node is a strong member of the community or not (an outlier). Such targets is exactly the scenario in which community-aware features should perform well. For empirical graphs described in Section~\ref{sec:real}, the target is a binary label that measures some practical feature or a role of a given node. 
It is important to highlight that these labels are not derived from the community structure of these graphs, at least not directly. Instead, they are characteristics of nodes defined independently of the graph structure. 
Therefore, for these networks we do not expect that community-aware features will significantly outperform other features. However, we conjecture that in many empirical networks, it may be the case that the prediction target is related to the fact that a node is a strong member of its own community or not. 
We expect to see that some community-aware features are still useful in prediction. It is important to highlight that, as we have described in Section~\ref{sec:real}, we have not hand-picked a few empirical networks that present good performance of community-aware features, aiming for a diverse collection of networks. 


\subsubsection*{Results and Observations}

In Tables~\ref{tab:IOX1} and~\ref{tab:IOX2}, we report the Kendall correlation for synthetic \textbf{ABCD+o} graphs and, respectively, empirical networks. In both tables, rows are sorted by the geometric mean across all investigated graphs so that features that provide more additional information are listed first.

\medskip

For artificial \textbf{ABCD+o} graphs, we observe the following patterns in Table~\ref{tab:IOX1}:
\begin{itemize}
    \item The lowest correlation is generally for measures related to a single community (\texttt{CADA}, \texttt{CADA*}, \texttt{CAS}), followed by measures taking into account all communities (\texttt{CPC} and the \texttt{CD\_} family of measures); in particular, \texttt{WMD} has the highest correlation with the classical features.
    \item The correlation decreases as the level of noise in the graphs increases.
\end{itemize}

The observed values are generally low which indicates that for artificial graphs, community-aware features are difficult to predict given classical graph features. The highest correlation value for \texttt{WMD} is not surprising since, in general, it correlates with the degree centrality.

\medskip

For empirical graphs, in Table~\ref{tab:IOX2} we observe slightly higher correlation values than for synthetic networks but the ordering of correlation values is similar to the previous results. Higher correlation values indicate that the community structure of empirical graphs is related to other structural characteristics, as opposed to synthetic \textbf{ABCD+o} graphs. Nevertheless, the correlation values are not too close to 1 anyway, so they are not entirely predictable from classical features. In particular, for the \textbf{Grid} graph, the correlation values are similar to artificial graphs (slightly above 0.2 for single-community measures).

In summary, the results confirm that the information encapsulated in community-aware measures cannot be recovered with high precision using classical features (even including embeddings). In the following experiments, we investigate if this extra information is useful for the node classification task.

\begin{table}
\centering
\caption{Information overlap between community-aware and classical features. The maximum of Kendall correlation between target and predictions on test data set for \textbf{ABCD+o} graphs.}\label{tab:IOX1}
\begin{tabular}{|l|llll|}
        \hline
        target & $\xi=0.3$ & $\xi=0.4$ & $\xi=0.5$ & $\xi=0.6$\\
        \hline
        \texttt{CADA} & 0.3305 & 0.2541 & 0.2292 & 0.1766 \\
        \texttt{CADA*} & 0.3613 & 0.2877 & 0.2772 & 0.1713 \\
        \texttt{CPC} & 0.3540 & 0.3568 & 0.3231 & 0.3106 \\
        \texttt{CAS} & 0.4205 & 0.3584 & 0.3138 & 0.2167 \\
        \texttt{CD\_L21} & 0.4539 & 0.4043 & 0.3823 & 0.3313 \\
        \texttt{CD\_L22} & 0.6265 & 0.5589 & 0.5009 & 0.4492 \\
        \texttt{CD\_L11} & 0.5935 & 0.5571 & 0.5834 & 0.5648 \\
        \texttt{CD\_L12} & 0.6503 & 0.5799 & 0.5464 & 0.5188 \\
        \texttt{CD\_KL1} & 0.6991 & 0.6411 & 0.5918 & 0.4929 \\
        \texttt{CD\_HD1} & 0.6809 & 0.6334 & 0.6170 & 0.5584 \\
        \texttt{CD\_KL2} & 0.7453 & 0.6602 & 0.6090 & 0.5471 \\
        \texttt{CD\_HD2} & 0.7546 & 0.7119 & 0.6815 & 0.6352 \\
        \texttt{WMD} & 0.7670 & 0.7288 & 0.6915 & 0.6387 \\
        \hline
\end{tabular}
\end{table}

\begin{table}
\centering
\caption{Information overlap between community-aware and classical features. The maximum of Kendall correlation between target and predictions on test data set for empirical graphs.}\label{tab:IOX2}
\begin{tabular}{|l|llllll|}
        \hline
        target & \textbf{Amazon} & \textbf{Facebook} & \textbf{Grid} & \textbf{LastFM} & \textbf{Reddit} & \textbf{Twitch} \\
        \hline
        \texttt{CADA} & 0.5830 & 0.5666 & 0.2156 & 0.4815 & 0.6826 & 0.5736 \\
        \texttt{CADA*} & 0.6058 & 0.5828 & 0.2174 & 0.5058 & 0.6867 & 0.5813 \\
        \texttt{CPC} & 0.6338 & 0.5992 & 0.2193 & 0.5175 & 0.7193 & 0.6219 \\
        \texttt{CAS} & 0.6538 & 0.6257 & 0.2999 & 0.5594 & 0.7306 & 0.6292 \\
        \texttt{CD\_L21} & 0.7052 & 0.6464 & 0.3496 & 0.5698 & 0.7574 & 0.6651 \\
        \texttt{CD\_L22} & 0.7554 & 0.7355 & 0.3557 & 0.6295 & 0.7941 & 0.6744 \\
        \texttt{CD\_L11} & 0.7251 & 0.7041 & 0.6978 & 0.6220 & 0.7735 & 0.6833 \\
        \texttt{CD\_L12} & 0.7794 & 0.7785 & 0.6447 & 0.6884 & 0.7810 & 0.7024 \\
        \texttt{CD\_KL1} & 0.7176 & 0.7516 & 0.7394 & 0.6289 & 0.7755 & 0.7087 \\
        \texttt{CD\_HD1} & 0.7383 & 0.7482 & 0.7168 & 0.6459 & 0.7853 & 0.7178 \\
        \texttt{CD\_KL2} & 0.7706 & 0.7826 & 0.7292 & 0.6853 & 0.8097 & 0.7405 \\
        \texttt{CD\_HD2} & 0.8212 & 0.8173 & 0.6930 & 0.7369 & 0.8221 & 0.7612 \\
        \texttt{WMD} & 0.8447 & 0.8456 & 0.8488 & 0.8531 & 0.7638 & 0.7414 \\
        \hline
\end{tabular}
\end{table}

\subsubsection{One-way Predictive Power}\label{sec:experiment2}

For the next experiment, for each graph each feature was considered individually as the only predictor except 
the two embeddings (\texttt{node2vec} and \texttt{struc2vec}) for which 16 dimensional vectors were taken as sets used to predict features.

With a 70/30 train-test split of the data (stratified by class labels since in some cases, the target feature is significantly unbalanced), a random forest model was built and two measures of predictive power are reported below: the area under the ROC curve (\emph{AUC}, computed using \texttt{roc\_auc\_score} function in \emph{scikit-learn})
and the average precision score (\emph{APS}, computed using \texttt{average\_precision\_score} function in \emph{scikit-learn}). Both measures were computed for a binary target node attribute as described in column ``target description'' in Table \ref{tab:real_networks} for real graphs and the outlier marker for the \textbf{ABCD+o} graphs.
We report two scores, since classes in the selected datasets are unbalanced. 
Indeed, in such cases the commonly used \emph{AUC} measure might not provide enough insight and \emph{APS} could be a better measure to pay attention to.
As a robustness check, we tried other prediction models (xgboost, lightgbm, regularized logistic regression) and obtained similar results (not reported in the paper but available on GitHub repository).


\subsubsection*{Results and Observations}

\begin{figure}[ht]
\centering
\includegraphics[width=0.47\textwidth]{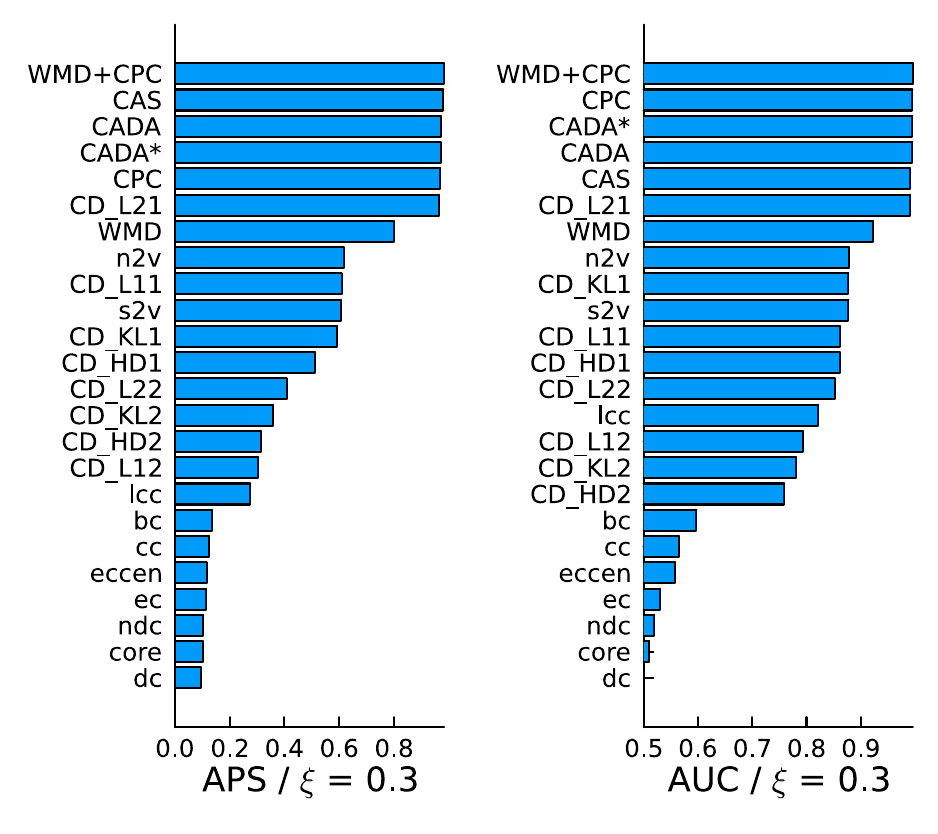}
\includegraphics[width=0.47\textwidth]{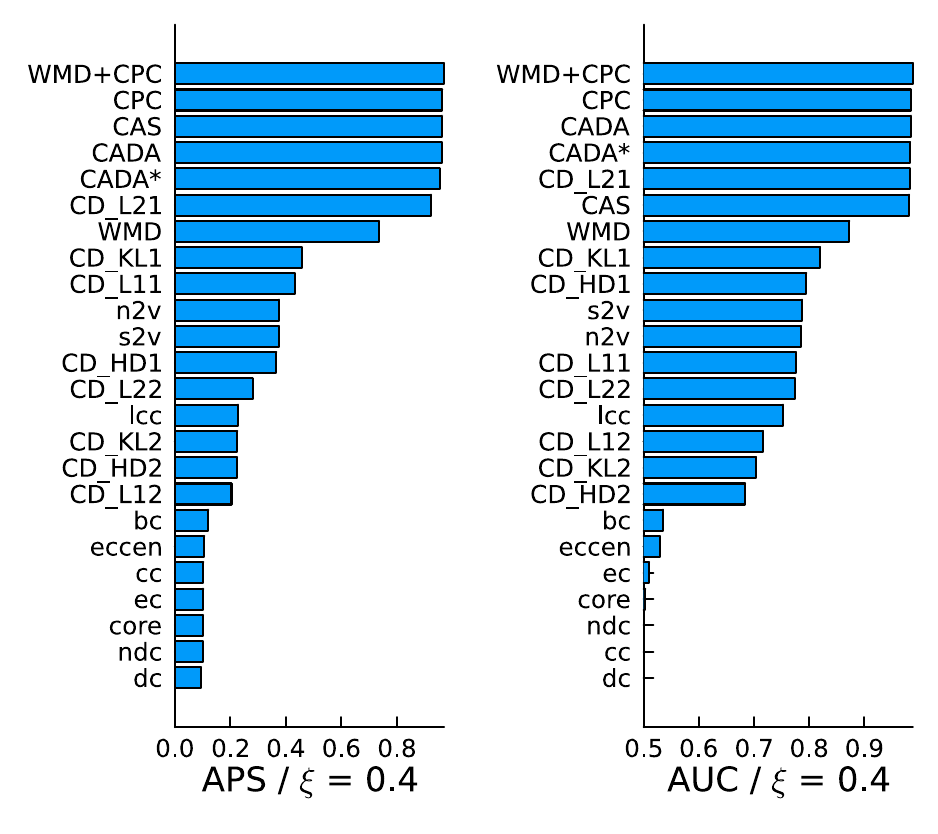}
\includegraphics[width=0.47\textwidth]{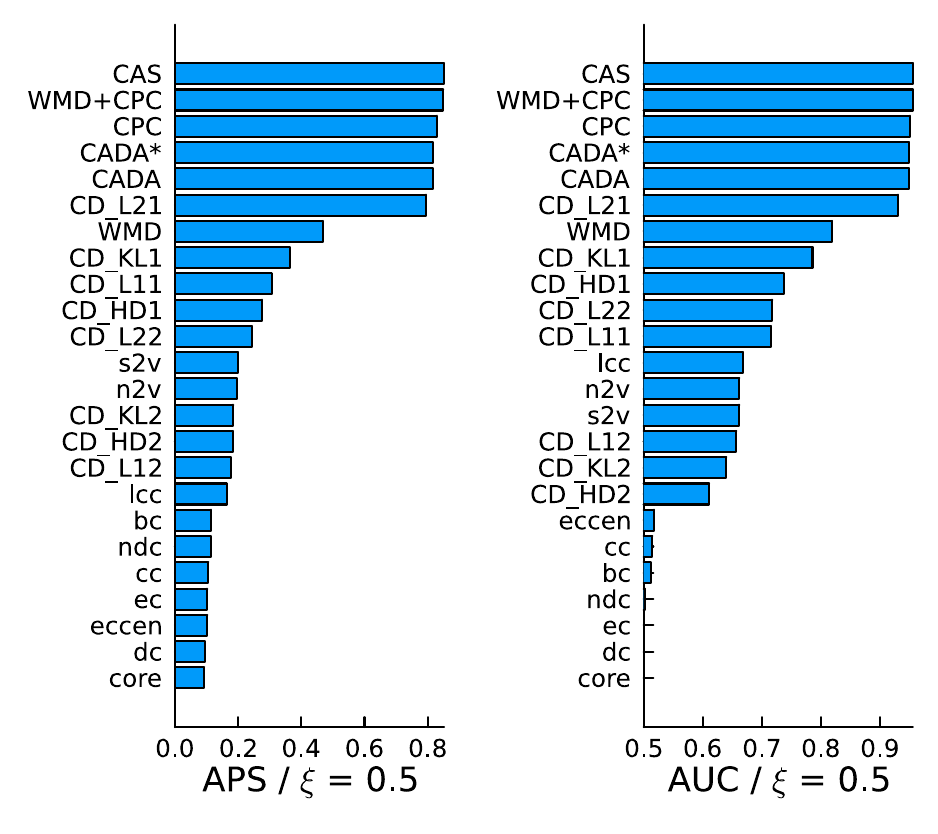}
\includegraphics[width=0.47\textwidth]{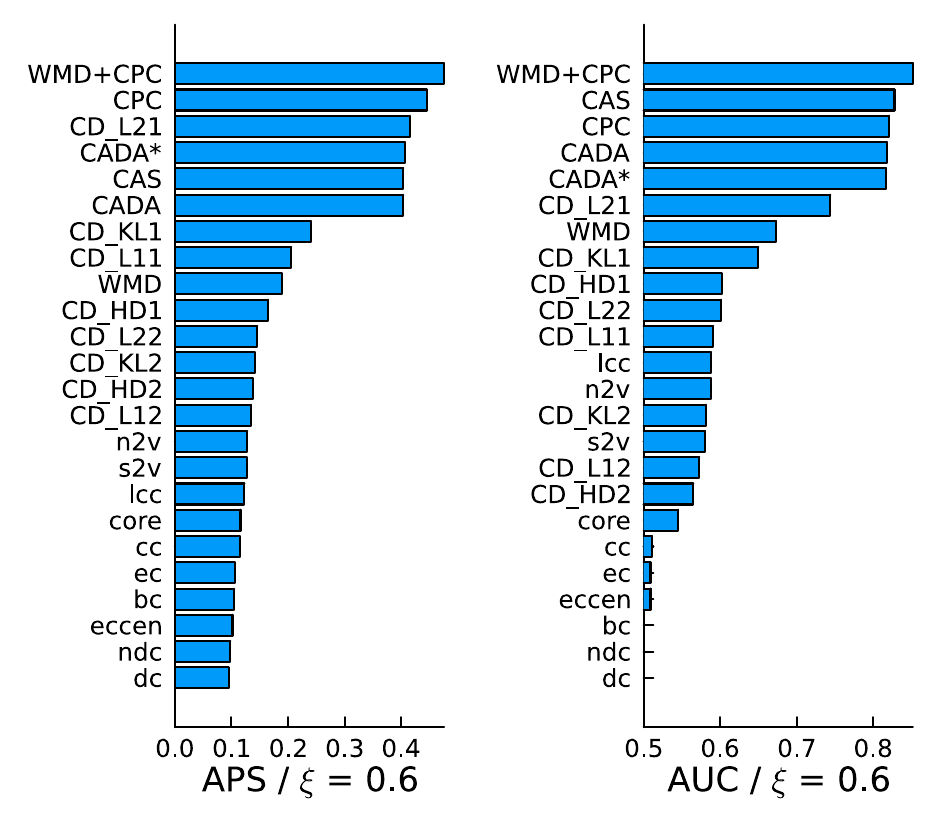}
\caption{Results of one-way predictive power assessment of considered node features for \textbf{ABCD+o} graphs}\label{fig:ow1}
\end{figure}

\begin{figure}[ht]
\centering
\includegraphics[width=0.46\textwidth]{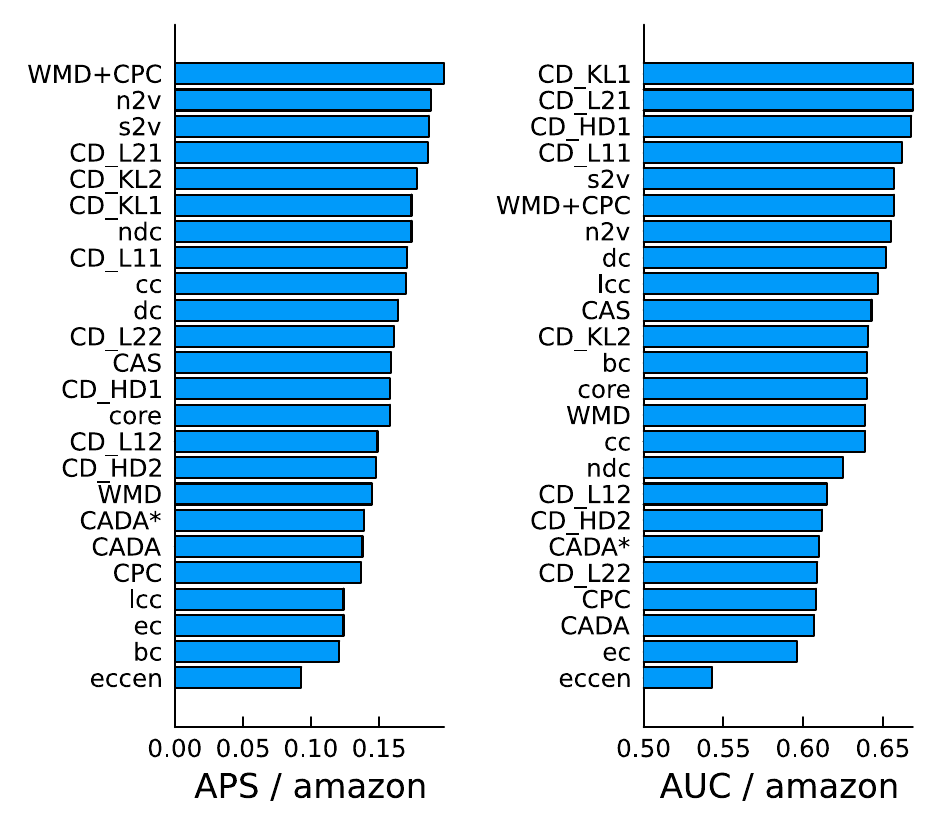}
\includegraphics[width=0.46\textwidth]{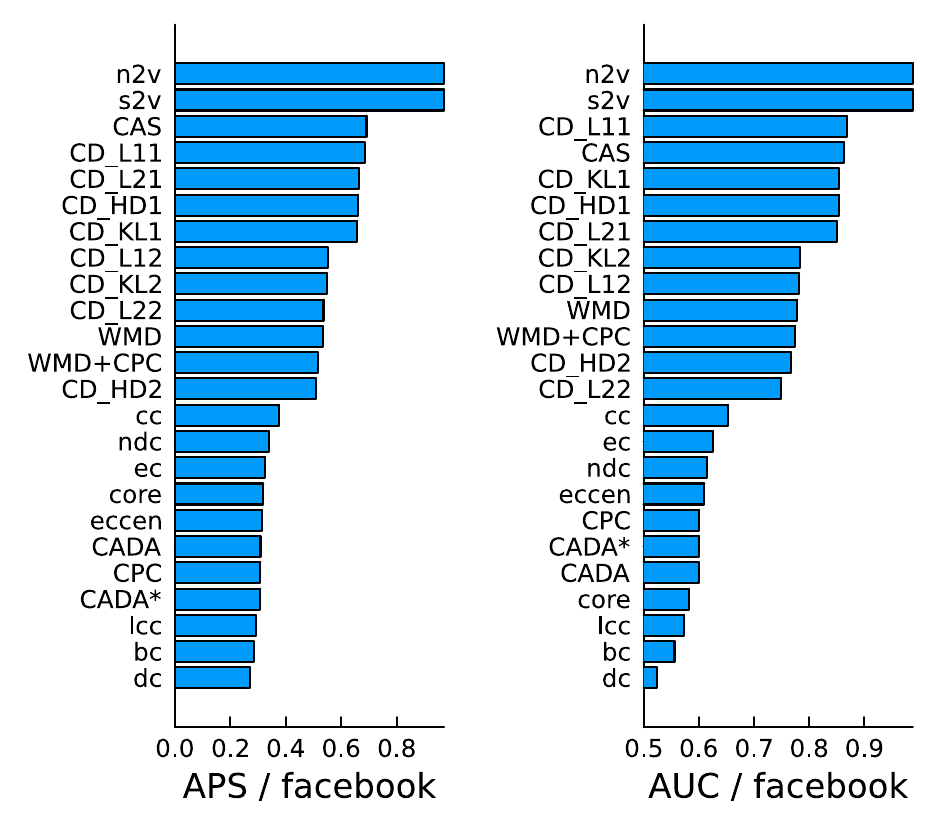}
\includegraphics[width=0.46\textwidth]{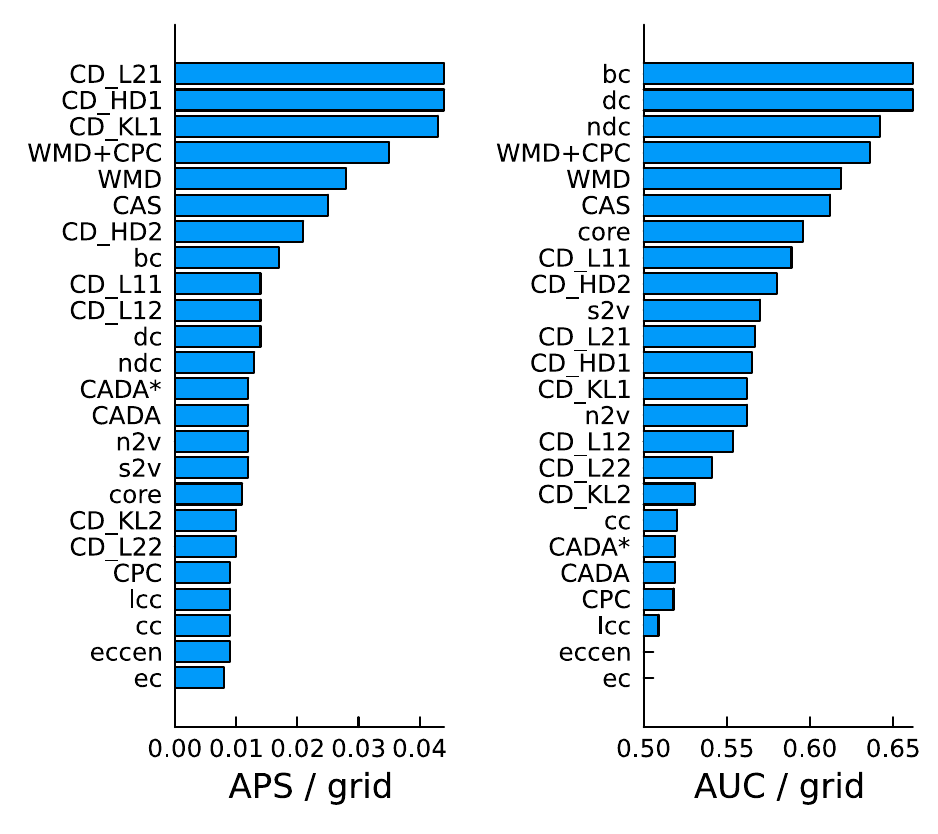}
\includegraphics[width=0.46\textwidth]{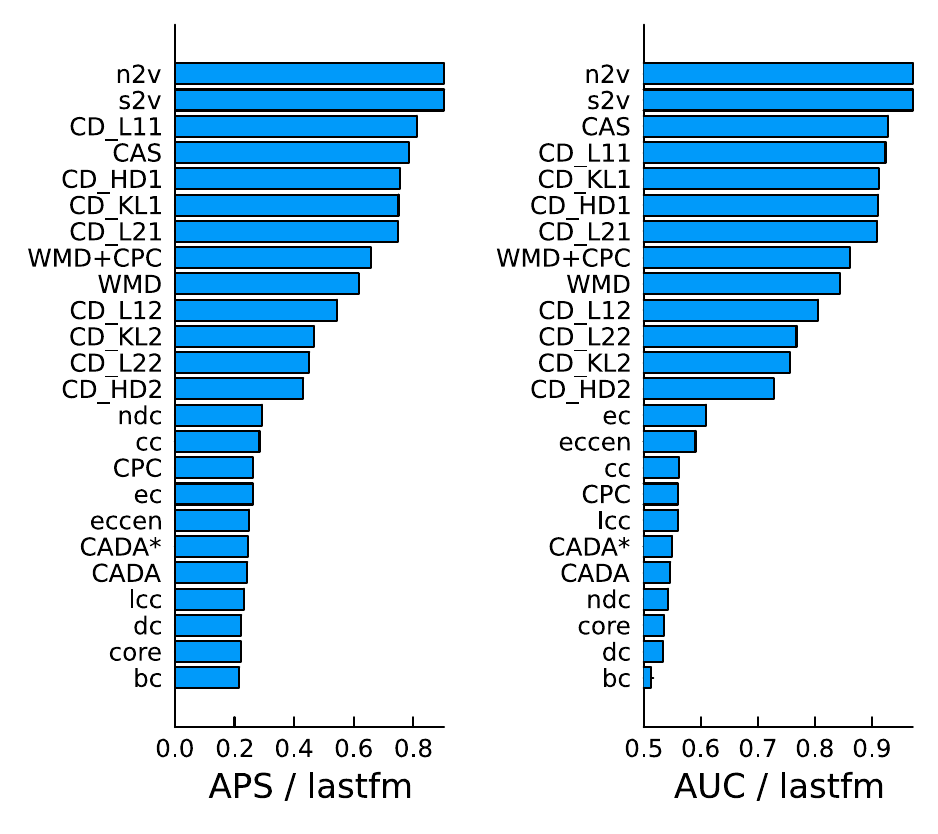}
\includegraphics[width=0.46\textwidth]{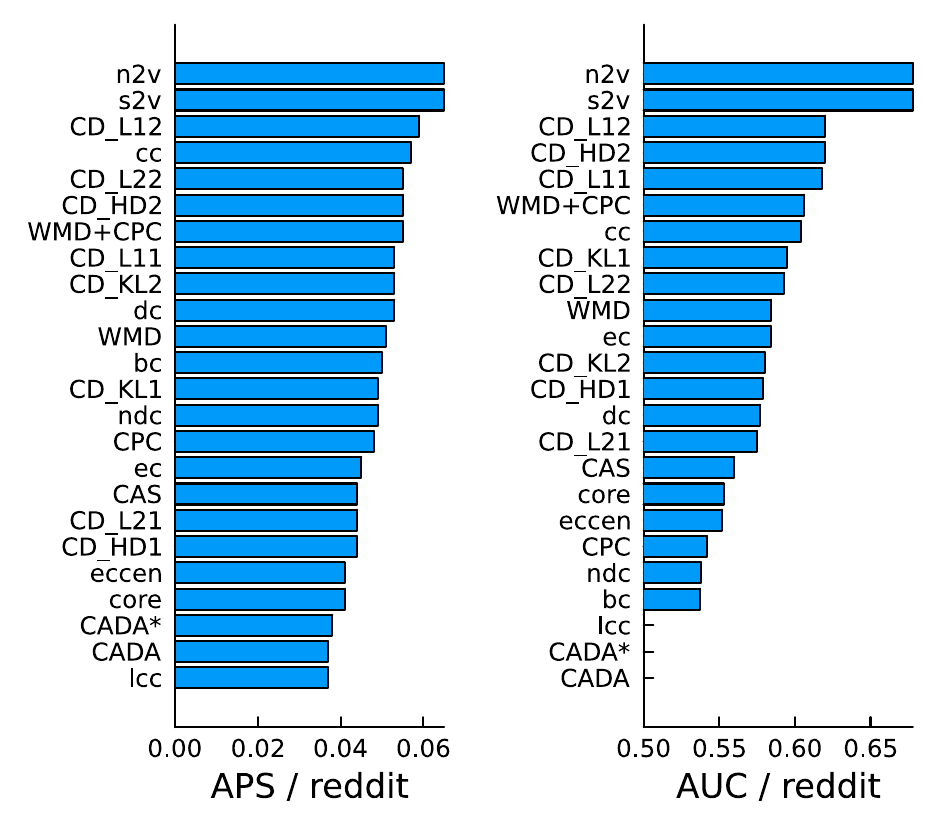}
\includegraphics[width=0.46\textwidth]{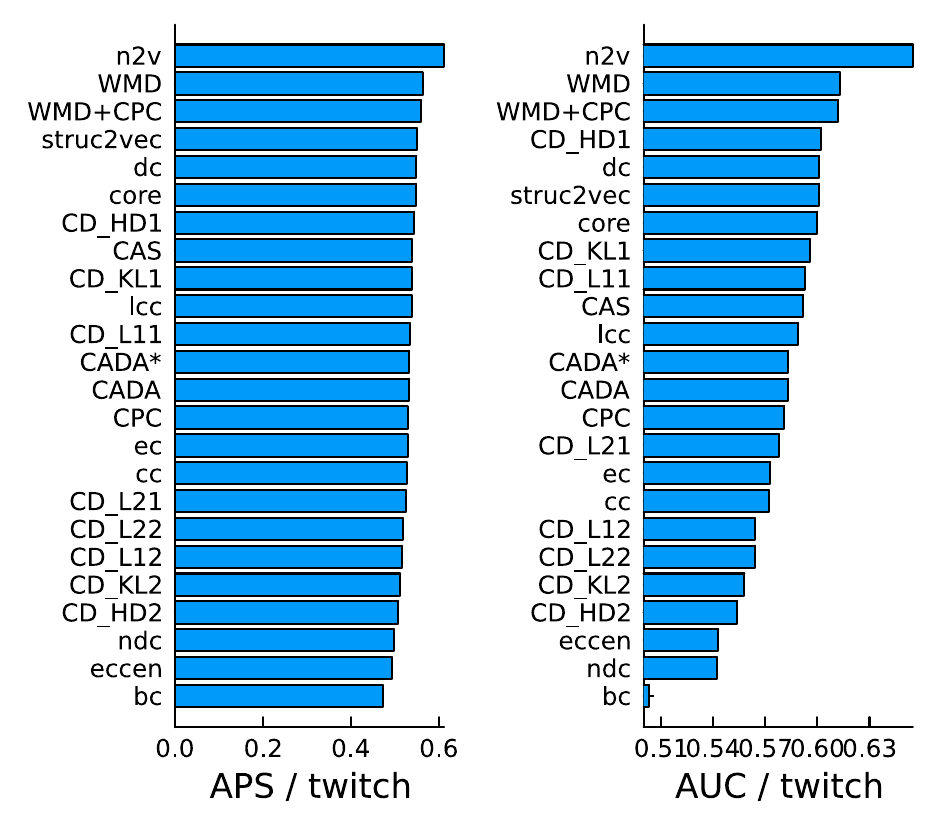}
\caption{Results of one-way predictive power assessment of considered node features for empirical graphs}\label{fig:ow2}
\end{figure}

Results of the experiments are reported in Figures~\ref{fig:ow1} and~\ref{fig:ow2} for each individual feature used as predictor.
Since for \texttt{node2vec} and \texttt{struc2vec} node embeddings all 16 dimensions are considered, we should expect better predictive power of such features and so better results. 
Similarly, as discussed in Section~\ref{sec:participation_score}, \texttt{WMD} and \texttt{CPC} are often considered together so we additionally consider 2-dimensional vectors consisting of these measures (\texttt{WMD+CPC}) as an input.
Both \emph{APS} and \emph{AUC} measures are reported. They are generally similar but not in all cases. In particular, we observe the largest difference for the \textbf{Grid} graph, which has the most imbalanced target variable.

For the \textbf{ABCD+o} graphs reported in Figure~\ref{fig:ow1}, the results indicate that community-aware features outperform classical features by a large margin for all levels of noise present in these synthetic networks.

For the empirical graphs reported in Figure~\ref{fig:ow2}, the results are more varied. For some graphs such as \textbf{Facebook}, \textbf{LastFM},  \textbf{Reddit}, and \textbf{Twitch}, the embeddings perform the best. However, in all cases the community-aware features are among the top scoring features and, in general, they score better than classical features excluding embeddings.
Recall that embeddings have the advantage that they are 16-dimensional while the community-aware features are just (1-dimensional) real numbers. As a result, embeddings are able to encapsulate more information about nodes and so they are expected to potentially be able to score higher.
It is also worth noting that \texttt{CAS} typically performs much better than \texttt{CADA} and \texttt{CADA*}, which indicates that taking into account the expected distribution of neighbours across communities based on the null-model gives additional and valuable information.

In summary, the one-way analysis confirms the usefulness of community-aware features both in synthetic as well as empirical graphs. This finding is consistent with the next computational experiment.

\clearpage

\subsubsection{Combined Variable Importance for Prediction}\label{sec:experiment3}

The third experiment (\emph{combined variable importance for prediction}) provides yet another way to verify the usefulness of community-aware features for node classification task. 
For each graph we take the same target as in the \emph{one-way predictive power} experiment, but this time we build a single model that takes into account all community-aware as well as all classical features (including both embeddings) as explanatory variables. A random forest classifier was built. For each variable, we computed its importance using the permutation approach described in~\cite{breiman2001,fisher2019models}. The variable importance was computed for each feature using \emph{APS} as a target predictive measure.

As in the previous experiments, a 70/30 train-test split was used.
We report the ranking of variable importance (rank 1 being the most important one) so that the values are comparable across all graphs investigated in this experiment. The raw importance scores have different ranges for various graphs.


\subsubsection*{Results and Observations}

The results are presented in Tables~\ref{tab:VI1} and~\ref{tab:VI2} for synthetic graphs and, respectively, empirical ones.
The ranks range between 1 and 53 (with rank 1 being the best), since there are 53 features in total (13 community-aware, 8 classical, 16 for \texttt{node2vec}, and 16 for \texttt{struc2vec}). The rows are sorted by the arithmetic mean of rank correlations across all graphs. We added the APS and AUC of the models used to derive the variable importance.

The observations are consistent with the results of the one-way predictive power experiment:
\begin{itemize}
    \item For \textbf{ABCD+o} graphs, the community-aware features perform better than classical features, and \texttt{CAS} is better than \texttt{CADA}/\texttt{CADA*}, which again shows that considering a null-model distribution of edges across communities is informative.
    \item For empirical graphs the situation is more interesting. For one of them (namely, the \emph{Facebook} graph), no community-aware measure appears in the top-10. It should be noted though, as can be seen in Figure~\ref{fig:ow2}, that both \texttt{node2vec} and \texttt{struc2vec} embeddings provide almost perfect prediction for this graph. On the other hand, for the \emph{Grid} graph, community-aware features are important (3 of them are in the top-10). In general, the community-aware features that score high for at least one graph are: \texttt{CAS}, \texttt{CD\_L22}, \texttt{WMD}, \texttt{CD\_L12}, \texttt{CD\_HD2}, \texttt{CD\_HD1}, and \texttt{CD\_KL1}. In particular, we see that the second-neighbourhood measures are well represented. This indicates that looking at the community structure of larger ego-nets of nodes is useful for empirical graphs. This was not the case for synthetic \textbf{ABCD+o} graphs as their generation structure is simpler than the more sophisticated mechanisms that lead to network formation of empirical social networks.
\end{itemize}

\begin{table}
\centering
\caption{Variable importance ranks for community-aware features in models including all features as explanatory variables for \textbf{ABCD+o} graphs. Values range from 1 (the best) to 53 (the worst), as well as the APC and AUC measures of the model quality on the test datasets.}
\vspace{.25cm}
\label{tab:VI1}
\begin{tabular}{|l|rrrr|}
        \hline
        variable & $\xi=0.3$ & $\xi=0.4$ & $\xi=0.5$ & $\xi=0.6$\\
        \hline
        \texttt{CAS} & 1 & 1 & 1 & 1 \\
        \texttt{CD\_L22} & 6 & 16 & 7 & 18 \\
        \texttt{CD\_KL1} & 12 & 7 & 10 & 15 \\
        \texttt{CADA} & 5 & 2 & 3 & 2 \\
        \texttt{CD\_L21} & 4 & 3 & 5 & 5 \\
        \texttt{WMD} & 7 & 6 & 6 & 4 \\
        \texttt{CADA*} & 3 & 4 & 2 & 6 \\
        \texttt{CPC} & 2 & 5 & 4 & 3 \\
        \texttt{CD\_L12} & 9 & 18 & 12 & 28 \\
        \texttt{CD\_KL2} & 8 & 9 & 9 & 42 \\
        \texttt{CD\_L11} & 14 & 12 & 35 & 14 \\
        \texttt{CD\_HD2} & 10 & 13 & 15 & 22 \\
        \texttt{CD\_HD1} & 16 & 11 & 33 & 40 \\
        \hline
        APS & 0.9883 & 0.9791 & 0.8743 & 0.5348 \\
        AUC & 0.9979 & 0.9934 & 0.962 & 0.8522 \\
        \hline
\end{tabular}
\end{table}

\begin{table}
\centering
\caption{Variable importance ranks for community-aware features in models including all features as explanatory variables for empirical graphs. Values range from 1 (the best) to 53 (the worst), as well as the APC and AUC measures of the model quality on the test datasets.}
\vspace{.25cm}
\label{tab:VI2}
\begin{tabular}{|l|rrrrrr|}
        \hline
        variable & \textbf{Amazon} & \textbf{Facebook} & \textbf{Grid} & \textbf{LastFM} & \textbf{Reddit} & \textbf{Twitch} \\
        \hline
\texttt{CAS} & 16 & 17 & 6 & 6 & 40 & 15 \\
\texttt{CD\_KL1} & 18 & 20 & 14 & 9 & 30 & 5 \\
\texttt{WMD} & 49 & 25 & 1 & 8 & 31 & 12 \\
\texttt{CD\_L21} & 19 & 32 & 11 & 29 & 25 & 26 \\
\texttt{CADA} & 26 & 33 & 22 & 15 & 33 & 25 \\
\texttt{CPC} & 39 & 30 & 24 & 17 & 26 & 20 \\
\texttt{CD\_L22} & 25 & 28 & 3 & 11 & 49 & 31 \\
\texttt{CADA*} &37 & 34 & 26 & 14 & 50 & 24 \\
\texttt{CD\_HD1} & 23 & 23 & 17 & 27 & 8 & 10 \\
\texttt{CD\_L12} & 24 & 53 & 20 & 7 & 4 & 34 \\
\texttt{CD\_L11} & 14 & 22 & 18 & 45 & 27 & 21 \\
\texttt{CD\_KL2} & 15 & 31 & 27 & 42 & 28 & 41 \\
\texttt{CD\_HD2} & 9 & 52 & 46 & 38 & 32 & 38 \\
\hline
        APS & 0.2395 & 0.9585 & 0.0475 & 0.8863 & 0.0883 & 0.6469 \\
        AUC & 0.7375 & 0.9832 & 0.6814 & 0.9679 & 0.6805 & 0.6823 \\
        \hline
\end{tabular}
\end{table}

\section{Concluding Remarks}

In summary, community-aware features are useful for prediction of labels of nodes. We confirmed this hypothesis on synthetic graphs in which community-aware features clearly outperformed other classical node features. For the experiments on empirical graphs, it is important to highlight that we did not hand pick graphs for which the community-aware features would work well, but rather defined \emph{a priori} criteria for graph selection. In this way, we believe that what is reported is a fair assessment of how community-aware features are expected to perform in practice. In the experiments on empirical graphs, community-aware features were not always the most important ones (but sometimes they were). In particular, as expected, node embeddings performed well. Nevertheless, we are convinced that, given the observed predictive power of the features in diverse empirical graphs, it is recommended to include them in predictive models. Indeed, for certain graph structure-target variable combinations they turned out to be essential for obtaining a good predictive model.

Moreover, it should be highlighted that community-aware features have a relatively low computational complexity compared to many classical features or node embeddings. Hence, for large graphs where other computations may be prohibitive, community-aware features are of even more value . As an example, consider the \textbf{Facebook} graph for which community-aware features performed relatively poorly. However, if this graph were much larger making it challenging to compute \texttt{node2vec} or \texttt{struc2vec} embeddings for it, Figure~\ref{fig:ow2} indicates that the next best features were \texttt{CAS} and some distribution-based community-aware features.

Another important and desired property of community-aware features is that they are easily interpretable. After building a predictive model, the analyst can more easily explain what indeed could be the underlying reason for the prediction. For example, for some prediction problems being a strong member of a community might be a positive information, while in other cases it could be the opposite. This explainability can be contrasted with embeddings that, although often having strong predictive power, do not help the user to understand the underlying reasons for the predictions.

Finally, let us note that the new measures proposed in this paper (that is, \texttt{CAS} and distribution-based ones), in general, performed better than community-aware features proposed earlier in the literature (namely, \texttt{CADA}, \texttt{CPC}, \texttt{WMD}). This shows that looking at how strongly a given node is a member of a community over what could be predicted by the null-model (that is, if the node is adjacent to randomly generated edges) is, indeed, an attractive approach that can be recommended to be used in practice.

\bibliography{ref}

\end{document}